\documentclass[onecolumn,10.5pt,compsoc]{Astro-new}

\usepackage[bf]{caption}
\usepackage{cuted}  
\usepackage{epstopdf}
\usepackage{enumitem}

\usepackage{graphicx}

\graphicspath{{Figures/}{figures/tem/}}
\DeclareCaptionLabelSeparator{threespace}{~\,~}
\usepackage{wrapfig}

\begin{document}

\pretitle{}
\title{Using fireball networks to track more frequent reentries: \textit{Falcon 9} upper-stage orbit determination from video recordings}
\author{Eloy Peña-Asensio$^{1,2}$\cor, Josep M. Trigo-Rodríguez$^{2,3}$, Marco Langbroek$^{4,5}$, Albert Rimola$^{1}$, and Antonio J. Robles$^{6}$}
\addr{1 Departament de Química, Universitat Autònoma de Barcelona (UAB), Bellaterra, Catalonia, Spain}
\addr{2 Institute of Space Sciences (CSIC), Cerdanyola del Vallès, Barcelona, Catalonia, Spain}
\addr{3 Institut d’Estudis Espacials de Catalunya (IEEC), Ed. Nexus, Barcelona, Catalonia, Spain.}
\addr{4 Leiden Observatory, Leiden University Faculty of Science, Leiden, the Netherlands}
\addr{5 Belgian Working Group Satellites (BWGS)}
\addr{6 Spanish Meteor Network (SPMN) team}

\coremail{\cor eloy.pena@uab.cat}

\abstracts{On February 16 of 2021, an artificial object moving slowly over the Mediterranean was recorded by the Spanish Meteor Network (SPMN). Based on astrometric measurements, we identified this event as the reentry engine burn of a \textit{SpaceX Falcon 9} launch vehicle’s upper stage. To study this event in detail, we adapted the plane intersection method for near-straight meteoroid trajectories to analyze the slow and curved orbits associated with artificial objects. To corroborate our results, we approximated the orbital elements of the upper stage using four pieces of "debris" cataloged by the US Government Combined Space Operations Center. Based on these calculations, we also estimated the possible deorbit hazard zone using the MSISE90 model atmosphere. We provide guidance regarding the interference that these artificial bolides may generate in fireball studies. Additionally, because artificial bolides will likely become more frequent in the future, we point out the new role that ground-based detection networks can play in the monitoring of potentially hazardous artificial objects in near-Earth space and in determining the strewn fields of artificial space debris.}


\section{Introduction}

\noindent Interplanetary meteoroids generate fireballs when penetrating the Earth’s atmosphere in a range of velocities between $11.2\,km/s$, which is the minimum velocity required for attraction by the Earth, and $73\, km/s$, which is the maximum velocity that a natural body gravitationally bound to our solar system can achieve \cite{ceplecha1998meteor}. These luminous phenomena are generated by large meteoroids on the centimeter or meter scale impacting the atmosphere at hypersonic velocities and ablating their components as they collide with air particles and heat up \cite{opik1959distribution, revelle1979quasi, trigo2019flux}. Some of these bodies undergo catastrophic disruption and disintegrate completely, whereas others survive atmospheric entry and are deposited on the Earth's surface. These surviving materials, if they are pristine rocks of natural origin, are called meteorites \cite{rubin2010meteorite}.

To gain a better understanding of these events, meteor detection networks have been developed worldwide to monitor the sky and obtain valuable information regarding the characteristics, fates, and origins of meteoroids \cite{jacchia1956harvard, ceplecha1957photographic, bland2004desert, trigo2005development, weryk2007southern, gritsevich2014first, colas2015french, colas2020fripon, gardiol2016prisma, devillepoix2020global}. Since 1995, the Spanish fireball and meteorite recovery network (SPMN) has been operating in Spain and is distributed throughout the peninsular and insular territory with 30 ground-based stations equipped with all-sky charge-coupled device cameras and wide-field video systems \cite{trigo2005development, trigo20072006, madiedo2007multi}. The SPMN records the sky on a full-time basis and automatically detects any moving luminous objects up to a magnitude of 10.

However, not all fireballs recorded by these detection systems have natural origins. Some luminous events are generated by artificial meteors created by human space programs and the proliferation of satellite technology. The ever-increasing number of new launches, collisions in space, and debris-shedding events generate space debris in low Earth orbits \cite{klinkrad2010space}. This technogenic pollution generates a two-fold problem: i) it poses a risk for space exploration and can cause damage when debris fall back to the ground \cite{liou2006risks}, and ii) it can interfere with astronomical observations \cite{hainaut2020impact}. In this regard, space debris in low orbits undergoes gradual decay induced by atmospheric drag until they eventually burn up or flare by reflecting sunlight in favorable geometries. These events are largely undesirably recorded by optical systems \cite{bagrov2010calculation, mironov2021retrospective}. The detection and analysis of such events can contribute to monitoring the deorbits of hazardous artificial objects in near-Earth space.

Although artificial object reentries are eventually analyzed through ground-based observations (e.g., the simple-return capsule \textit{Genesis} \cite{revelle2005genesis}, robotic space probe \textit{Stardust} \cite{revelle2007stardust, levit2008reconstruction}, cargo spacecraft \textit{Jules Verne ATC} \cite{pas2009atv}, and asteroid explorer \textit{Hayabusa} \cite{ueda2011trajectory, shoemaker2013trajectory}), no meteor detection networks have developed and systematically implemented detection and reduction algorithms specifically for artificial meteor analysis.

Given the increasing number of such events and the need to discern their origins and fates, we adapted the SPMN network reduction method, which is available in the new \textit{3D FireTOC} software \cite{pena2021accurate}, for the automatic detection and analysis of objects with slow and curved trajectories. We tested our implementation through the study of the \textit{Falcon 9} reentry and compared the results to calculations based on debris orbital elements. Finally, we estimated the possible deorbit hazard zone using the MSISE90 atmospheric model.

\section{Dataset analyses}
\begin{figure}[htb]
\centering
\includegraphics[width=\textwidth]{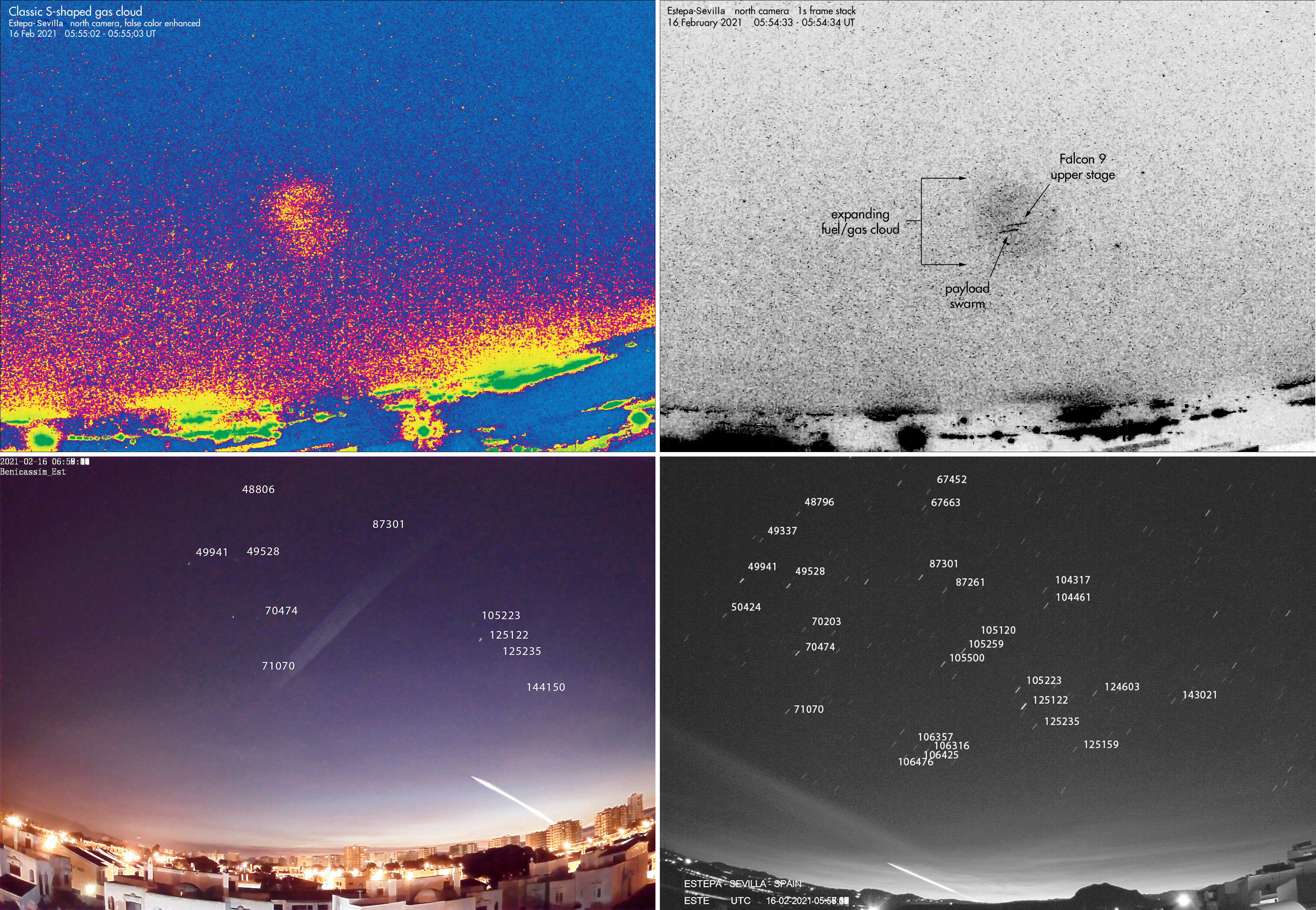}
\caption{Top left: classic S-shaped cloud from the Estepa North video with false color enhanced. Top right: \textit{Falcon 9} upper stage, payload swarm, and expanding fuel/gas cloud from Estepa North. Bottom left: overlaid frames and reference stars from Benicàssim. Bottom right: overlaid frames and reference stars from Estepa East.}
\label{fig:Fig1}
\end{figure}

\noindent We present two analyses of the event labelled SPMN160221ART recorded by the SPMN on February 16 of 2021, which crossed southern France in the direction of Libya. The fireball was captured by the north and east cameras of the Estepa station in Seville province, as well as the eastern camera of Benicàssim station in Castellón province (see Table \ref{table:Stations}). The object could be observed in the field of view for a minute and a half, which indicated a low velocity and elliptical or nearly circular orbit. As will be demonstrated in this paper, this event was generated by the upper stage and payloads of a \textit{Falcon 9} rocket that launched a batch of \textit{SpaceX Starlink} satellites into Earth orbit earlier that night. This rocket stage was deliberately deorbited over the Indian Ocean 1.5 revolutions (2.5 h) after launch (just after our observations). Figure \ref{fig:Fig1} presents enhanced images extracted from each video recording. When comparing the images to those captured by the International Space Station and images of \textit{Iridium} flares, we concluded that the artificial fireball exhibited an apparent magnitude of $-6\pm1$ at both monitoring stations.

\begin{table}[htbp]
\vspace {-3mm}
  \caption{SPMN stations recording the SPMN160221ART event on 2/16/21.}
\vspace {-3mm}
\begin{center}
 \begin{tabular}{cccccc} 
 \hline
Station &	Longitude &	Latitude &	Alt. &	Start Time & End Time   \\ [0.5ex] 
 \hline 
 Estepa N &	4º 52´ 36" W &	37º 17´ 29" N & 	537 m & 05:53:36 UTC &05:56:30 UTC \\
Estepa E &	4º 52´ 36" W &	37º 17´ 29" N  & 	537 m &	05:55:53 UTC &	05:58:51 UTC\\
Benicàssim &	0º 02´ 19" E &	40º 02´ 03" N &	15 m &	05:57:38 UTC &	05:59:03 UTC\\ [0.5ex]
 \hline
\end{tabular}
  \label{table:Stations}
\end{center}
\end{table}

\subsection{Astrometric calibration}
\begin{figure}[htb]
\centering
\includegraphics[width=0.65\textwidth]{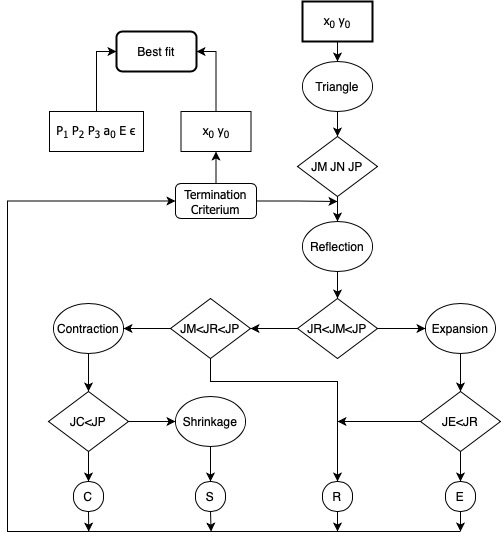}
\caption{Block diagram of the simplex method applied to the astrometry. $R$ is the substitution point for reflection, $E$ is the expansion, $C$ is the contraction, and $S$ is the shrinkage.}
\label{fig:Fig2}
\end{figure}

To reduce both recordings astrometrically, the first step is to perform astrometry on the stars recorded in the images containing the artificial object/bolide to calculate its projection onto the celestial sphere (i.e., its apparent trajectory) \cite{pena2021accurate}. By identifying the stars in the visible sky, a method can be applied to consider the distortion of the lens to determine the relationships between pixels and horizontal coordinates. Various calibration methods have been proposed for all-sky cameras \cite{ceplecha1987geometric, borovivcka1992astrometry, borovicka1995new}. These calibrations involve solving highly nonlinear equations such that convergence is nontrivial. Therefore, we implemented the polynomial variant proposed by \cite{bannister2013numerical}, which significantly improves the convergence of solutions. However, based on the low number of stars visible in the videos and unknown camera constants, we applied the simplex algorithm to optimize the initial values and guarantee a robust solution \cite{motzkin1956assignment}. The proposed method follows the diagram presented in Figure \ref{fig:Fig2}, assumes symmetrical lens distortion, and requires the determination of the parameters $P_1$, $P_2$, $P_3$, $a_0$, $E$, $\epsilon$, $x_0$, and $y_0$ according to the following equations:

\begin{equation}
r = \sqrt{(x-x_0)^2+(y-y_0)^2},
\end{equation}

\begin{equation}
u = P_1 r^2 + P_2 r + P_3,
\end{equation}

\begin{equation}
b = a_0 - E +  \tan^{-1} \left( \frac{y-y_0}{x-x_0}  \right),
\end{equation}

\begin{equation}
\cos(z) = \cos(u)\cos(\epsilon)-\sin(u)\sin(\epsilon)\cos(b),
\end{equation}

\begin{equation}
\sin(a-E) = \sin(b)\sin(u)/\sin(z),
\end{equation}
where $x_0, y_0$ is the center of projection (COP), where the system’s optical axis intersects the sensor plane; $r$, $u$, and $b$ are the radial distance, zenith-like angle mapping, and azimuth-like angle of a pixel coordinate and the COP, respectively; $a_0$ is the rotation of the sensor’s x axis from the cardinal south; $E$ is the rotation between the x axis and a vector defined by the true zenith projection and the COP; $\epsilon$ is the angle between the true zenith and COP; and $z$ and $a$ are the zenith angle and azimuth of the given pixel coordinate, respectively.

Once the cameras are calibrated, it is possible to transform pixel coordinates into horizontal coordinates and then equatorial coordinates. Based on the projections of the apparent trajectory onto the celestial sphere, the real trajectory can be reconstructed. However, because the time span of the video is relatively long, the rotational motion of the Earth is relevant and must be considered when transforming between coordinate systems.

\subsection{Orbit reconstruction from ground-based observations}
We originally implemented the triangulation method based on the intersection of planes proposed in \cite{ceplecha1987geometric}. However, to study this event, we had to adapt this method to compute curved paths such as satellite orbits because our implemented meteor analysis software was specifically designed to compute typical near-straight trajectories of meteoroids \cite{pena2020}. 

\begin{figure}[htb]
\centering
\includegraphics[width=0.75\textwidth]{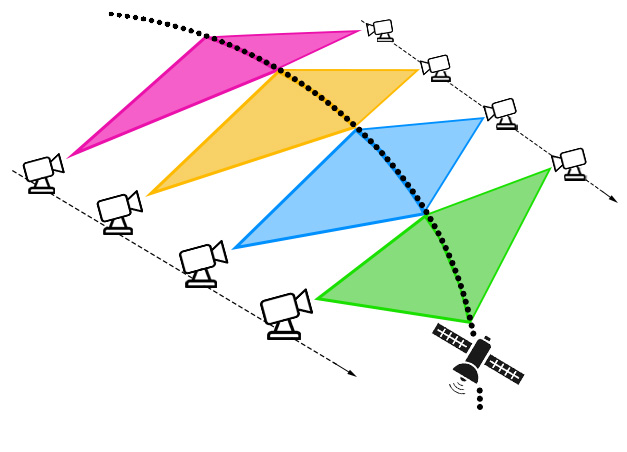}
\caption{Schematic diagram of the plane intersection method divided into segments to reconstruct curved trajectories considering the rotational motion of the Earth.}
\label{fig:Fig3}
\end{figure}

\begin{figure}[htb]
\centering
\includegraphics[width=0.9\textwidth]{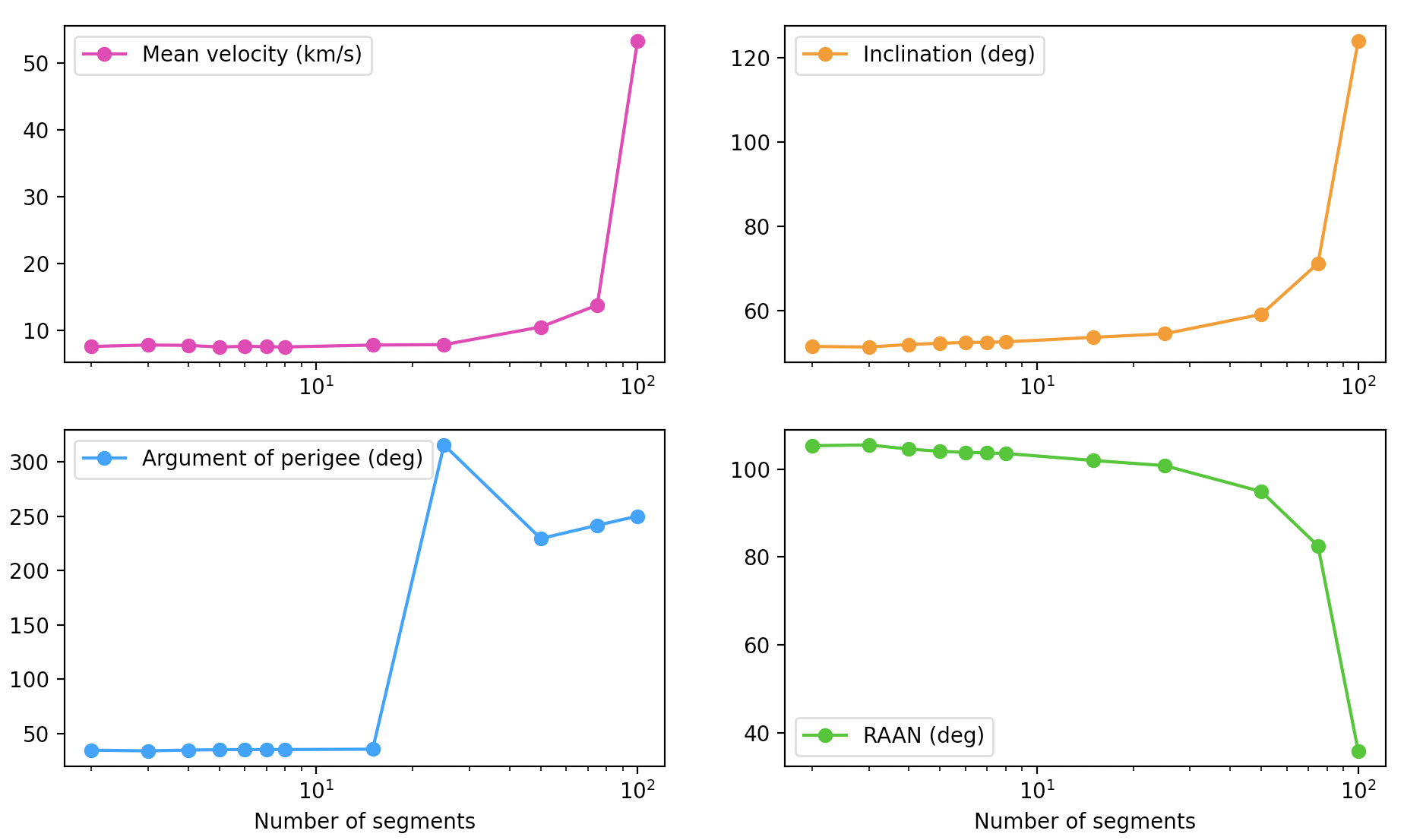}
\caption{Variation in mean velocity, inclination, argument of perigee, and right ascension of the ascending node (RAAN) as a function of the number of segments used in the modified plane intersection method.}
\label{fig:FigSegments}
\end{figure}

The mean plane containing the apparent trajectory of each station was obtained from the observation of a fireball from two or more stations. The intersection of these planes represents the atmospheric trajectory of the meteoroid. To reconstruct a curved orbit, we divided the observed trajectory into small segments such that each segment could better conform to a linear assumption (see Figure \ref{fig:Fig3}). In this manner, we obtained small straight sections that formed the curved trajectory when combined. The division into small segments was performed in accordance with the total number of observed points, which corresponded to an equally distributed duration because during the observation period, the velocity change was sufficiently slow (less than the uncertainty of the observed velocity). Because the plane intersection method corrects for point spread by calculating the mean plane of a path, overly small segments would be unable to compensate for deviations, whereas overly large segments would not correctly model the curved trajectory. Therefore, given the sensitivity of the measurements, by reducing the size of each segment and increasing the number of divisions, the triangulation method will begin to diverge at some point. As shown in Figure \ref{fig:FigSegments}, the results are robust below eight subdivisions. 

\begin{figure}[htb]
\centering
\includegraphics[width=0.65\textwidth]{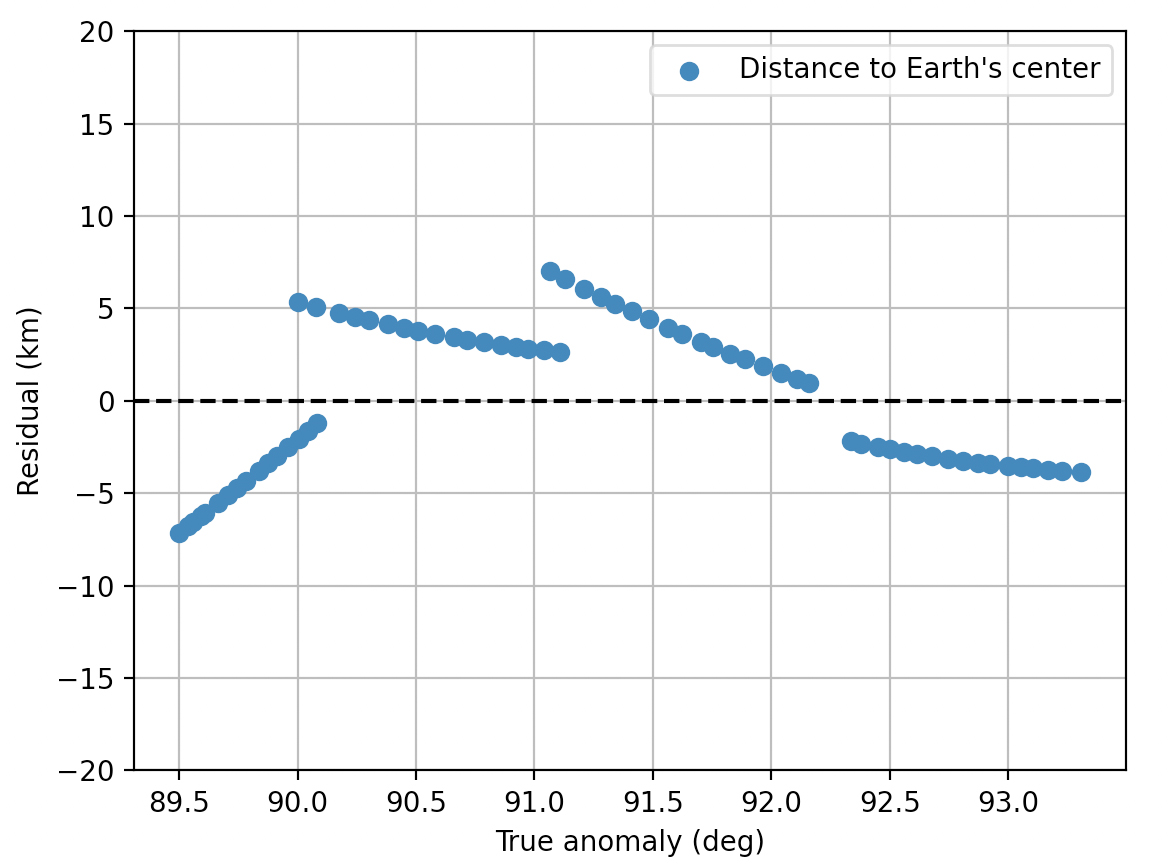}
\caption{Residuals of the elliptical orbit fitted by subdividing the observed path into four segments.}
\label{fig:FigResiduals}
\end{figure}

By fitting the mean plane containing the observed points (the plane to which they are the least distant on average), we calculated the plane of the orbit. Points that did not lie on this plane were projected perpendicularly onto the plane to minimize error. In this manner, we derived the corresponding Cartesian coordinates of each detected point such that the orbital state vector at a given epoch could be trivially derived by subtracting two consecutive positions. We achieved the best fit with four segments, resulting in minimized mean residuals. Figure \ref{fig:FigResiduals} presents the residuals of the fitted elliptical orbit.

The last step was to transform the state vector into orbital elements. Once we had the position and velocity vectors, we calculated the specific angular momentum $\bar{h}$ and node vector $\hat{n}$ as follows:

\begin{equation} \label{eq3}
\bar{h} = \bar{r} \times \bar{v},
\end{equation}
\begin{equation}
\bar{n} = (-h_y, h_x, 0).
\end{equation}

Then, all the orbital elements could be computed as
\begin{equation}
 a = \frac{1}{\frac{2}{r}- \frac{v^2}{GM}},
\end{equation}
\begin{equation}
 \bar{e}  = \frac{1}{GM} \left ( \left(v^2-\frac{GM}{r} \right )\bar{r}-(\bar{r}\cdot \bar{v}) \bar{v} \right ),
\end{equation}
\begin{equation} \label{eq4}
 i  = \arccos \left(  \frac{h_z}{h} \right ),
\end{equation}

\begin{equation}
 \Omega  = \arccos  \left( \frac{h_y}{\sqrt{h_x^2 + h_y^2}} \right ),
\end{equation}
\begin{equation}
 \omega  = \arccos \left( \frac{-h_y e_x + h_x e_y}{e\sqrt{h_x^2 + h_y^2}} \right ),
\end{equation}

\begin{equation}
 v_0  = \arccos \left( \frac{\bar{e}\cdot \bar{r}}{er} \right ),
\end{equation}

where $a$ is the semi-major axis, $e$ is the eccentricity, $i$ is the inclination, $\Omega$ is the longitude of the ascending node, $\omega$ is the argument of the perihelion, and $v_0$ is the true anomaly \cite{dubiago1961determination}.

\subsection{Proxy orbit computation from debris piece tracking}
After identifying the event as potentially being related to the \textit{Starlink V1.0-L19} launch on the same night, we obtained orbital elements for the objects from the US government’s Combined Space Operations Center (CSpOC) through their web portal \textit{Space-Track}. For this launch, 64 objects were cataloged: 60 payloads and four pieces of debris. Table \ref{table:Debris} lists the IDs of the four debris objects considered. The pieces of debris are four containment rods that were used to keep the payloads stacked on the \textit{Falcon 9} upper stage. They were jettisoned during payload release.

\begin{table}[htbp]
\vspace {-3mm}
  \caption{NORAD and COSPAR IDs of the debris pieces considered.}
\vspace {-3mm}
\begin{center}
 \begin{tabular}{ccc} 
 \hline
Debris &	NORAD ID &	COSPAR ID \\ [0.5ex] 
 \hline 
 1 &	47683 &	2021-012BR  \\
2 &	47682 &	2021-012BQ  \\
3 &	47681 &	2021-012BP\\
4 & 47680 & 2021-012BN\\
[0.5ex]
 \hline
\end{tabular}
  \label{table:Debris}
\end{center}
\end{table}

Because objects that make less than two revolutions before reentry are typically not cataloged, there are no tracking-based orbital elements for the \textit{Falcon 9} upper stage, which is the prime suspect for the event we observed. However, we can derive approximate orbital elements for the upper stage by using the four pieces of debris as a proxy. The debris were jettisoned upon payload release under the effects of inertia and subjected solely to drag (unlike the payloads, which subsequently maneuvered to higher orbits using their propulsion systems). The orbit of the \textit{Falcon 9} upper stage should initially closely match those of the four debris pieces. The first elements available for the four debris pieces were captured on February 23 of 2021, which was one week after launch. By using the SGP4 computational model \cite{Hoots1980, Vallado2006}, we propagated the orbits back to the moment at which they separated from the upper stage and satellite stack (February 16, 2021, 04:08:24 UTC). We then took the average of the four resulting orbital element sets as the first proxy for the orbit of the \textit{Falcon 9} upper stage. For the two observation stations at Estepa and Benicàssim, this first proxy orbit resulted in sky trajectories that closely match the observations with a small time difference $\Delta t$ of $\sim 17$ s for a given point on the sky trajectory. Small tweaks to the mean motion, eccentricity, and inclination were then made to reduce $\Delta t$ to near zero, resulting in a new final proxy orbit. 

We employed the \textit{SatFit 3.1} orbit fitting software written by Scott Campbell for this process\footnote{The source code is available at http://sat.belastro.net/satelliteorbitdetermination.com/}. \textit{SatFit} modifies the SGP4 orbital elements by using a least-squares fitting procedure to improve the fit of elements to astrometric observations. \textit{SatFit} provides feedback regarding the resulting fit by returning information on the overall positional error, cross-track error, and delta t. The mean motion, which represents the time it takes the rocket stage to complete one revolution around the Earth, was adjusted to a value yielding a $\Delta t$ of less than 1 s via fitting to four astrometry points from the early portion of the Estepa video. The inclination was adjusted to eliminate a small cross-track error. The adjustments amounted to $-0.00009242\, rev/day$ for mean motion and $0.08\, ^{\circ}$ for inclination. Small adjustments were also made in the RAAN ($-0.3029\, ^{\circ}$), eccentricity ($-0.0000026$), and mean anomaly to improve the $\Delta t$ fit and reduce the cross-track error (adjusting the inclination automatically means the RAAN must also be adjusted and adjusting the mean motion means the mean anomaly must also be adjusted).

\subsection{Reentry trajectory prediction}

The triangulation of the bright phase in the images from Estepa and Benicàssim facilitated the construction of a state vector and from this vector, a second set of orbital elements was derived (Table \ref{table:TLE}). The state vector was transformed into orbital elements using the \textit{RV2TLE} software written by Scott Campbell, yielding a set of SGP4-compatible orbital elements in the three-line-ephemerid (TLE) format\footnote{This format is commonly used for satellite orbits (for a description of the TLE format, see http://www.satobs.org/element.html)}.

\begin{figure}[htb]
\centering
\includegraphics[width=\textwidth]{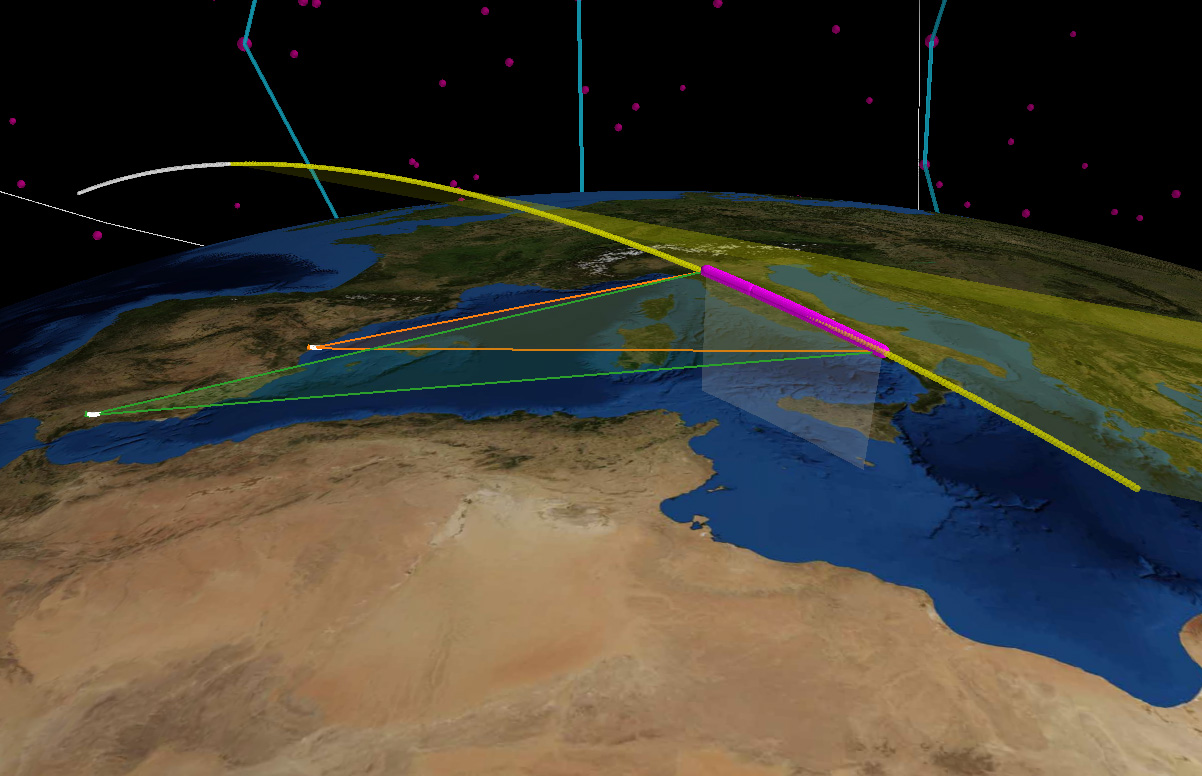}
\caption{3D scale representation of the \textit{Falcon 9} trajectory orbit. The proxy orbit is white (yellow when illuminated by sunlight), the observed path used for reduction is purple, the Estepa station is green, and the Benicàssim station is orange.}
\label{fig:Fig4}
\end{figure}

The proxy orbit of the \textit{Falcon 9} upper stage is consistent with the reduction performed on the ground-based video data, as can be observed in Figure \ref{fig:Fig4}, particularly in the orbital plane, but also in the orbital altitude.

\begin{table}[htbp]
\vspace {-3mm}
  \caption{TLE computed from the triangulation of SPMN data.}
 \vspace {-3mm}

 \begin{center}
 \begin{tabular}{l c} 
 \hline
Falcon 9 R/B (from State Vector)  \\ [0.5ex] 
 \hline
1{ }99999U{ }21012BS{ }{ }21047.24849537{ }0.00000000{ }{ }00000-0{ }{ }00000+0 0{ }{ }{ }{ }09\\ 
2{ }99999{ }{ }{ }{ }52.4033{ }104.0553{ }0228149{ }225.2325{ }{ }266.6565{ }16.05238701{ }{ }{ }{ }08\\ [0.5ex] 
 \hline

\end{tabular}
  \label{table:TLE}
 
\end{center}
\end{table}


Our triangulation calculations yielded an altitude of approximately $270\pm0.6\, km$ and velocity of $7.5\pm0.3\, km/s$ at 05:57:58 UTC. It should be noted that our observations likely captured the rocket stage just after it performed a deorbit burn. Therefore, small discrepancies are expected between our first orbital element set, the proxy orbit (which is the pre-burn orbit), and real observed trajectory. However, at this stage of the launch, the payloads should still be very close to the pre-burn proxy orbit for the upper stage. The second orbital element set derived from the state vector represents the post-burn reentry trajectory. This orbit has a semi-major axis of $6638\, km$ with a nominal apogee of $411\, km$ and perigee of $108\, km$. These are values with respect to the Earth’s equatorial radius. The real altitude depends on the location of the perigee (for this particular orbital revolution, the perigee is near $118\,km$ above the geoid). The orbital inclination is $52.4\, ^{\circ}$, which is a difference of a few tenths of a degree (hundred arcseconds) relative to the pre-burn proxy orbit discussed above. With an eccentricity of $0.0228$, this orbit is much more eccentric than the pre-burn proxy, as expected for a post-burn reentry trajectory.

\begin{figure}[htb]
\centering
\includegraphics[width=\textwidth]{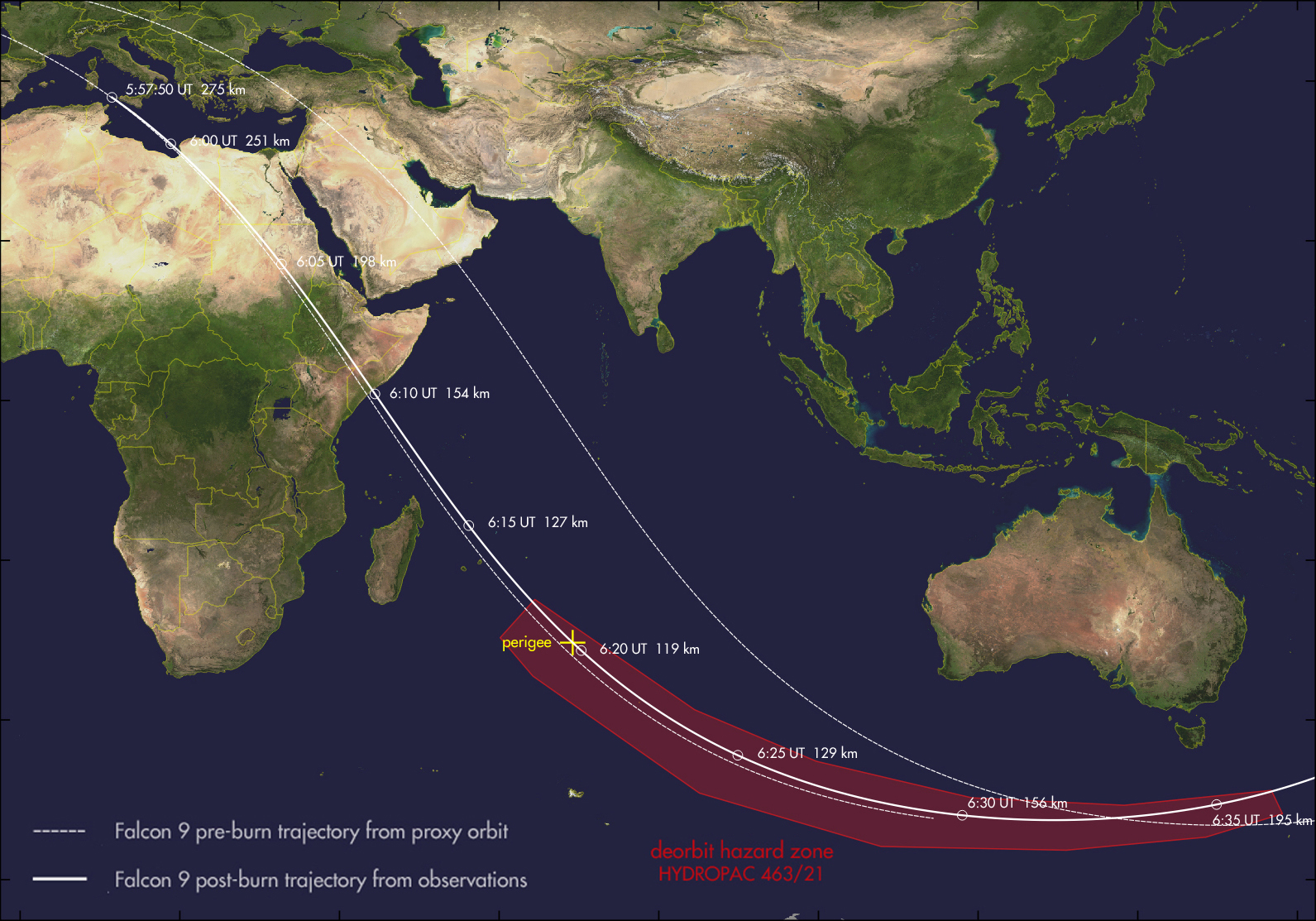}
\caption{Reentry trajectory estimation. The proxy orbit is the dashed white line and the triangulated orbit is the solid white line. The yellow cross marks perigee of this orbit. The area indicated in red is the deorbit hazard area identified by the \textit{Navigational Warning HYDROPAC 463/21.}}
\label{fig:Fig5}
\end{figure}

The perigee on this revolution is reached at approximately 6:19 UTC and is located near $30.1^{\circ}\, S$, $68.9^{\circ}\, E$. This is within the western portion of the deorbit hazard zone defined by \textit{Navigational Warning HYDROPAC 463/21}\footnote{ Source: https://msi.nga.mil/NavWarnings}. The map in Figure \ref{fig:Fig5} compares the orbit derived from the state vector (solid white line) to the pre-burn proxy orbit (thin dashed line). The yellow cross represents the perigee. Times are in UTC.

Compared to the pre-deorbit-burn proxy orbit based on the orbits of the four retaining rods, the post-deorbit-burn orbit from the triangulation-based state vector has a difference of $-0.732\, deg$ in inclination, $-11.28\, km$ in the semi-major axis, and $0.020423$ in eccentricity. 
The eccentricity of the post-reentry burn orbit must be larger than the eccentricity of the pre-deorbit-burn proxy orbit by definition. Its semi-major axis must also be smaller by definition.

The 3D positional difference between the two orbits is minimized ($10\, km$) at approximately 5:57:48 UTC. The cross-plane difference is minimized ($5\, km$) 45 s earlier at approximately 5:56:58 UTC. At the state vector epoch (05:57:50 UTC), the absolute positional difference between the pre-deorbit-burn proxy orbit and state vector is approximately $10.2\, km$, $1.8\, km$ of which is in altitude and $8.4\, km$ is in the cross-plane (horizontal) direction. Compared to the inherent positional accuracies of SGP4 (~$1\, km$ at epoch and increasingly more before and after that), these differences are small and reasonable differences indicating a good fit between the proxy orbit and orbit derived from triangulation \cite{osweiler2006covariance, kelso2007validation}. 

The moment of the smallest positional difference (5:57:48 UTC) is very close to the state vector epoch, which is close to the moment at which the trail on the images noticeably brightened. This may indicate that this moment captures the start of the actual deorbit burn. Because the deorbit burn has a duration, this could indicate that the state vector we obtained still underestimated the eccentricity and overestimated the perigee altitude of the final reentry orbit.

The nominal perigee altitude derived from the state vector appears to be slightly too high. A reentry model was simulated using the \textit{NASA General Mission Analysis Tool (GMAT) R2020a} software \cite{Moriba2009}\footnote{GMAT is downloadable at https://sourceforge.net/projects/gmat/} with the MSISE90 model atmosphere, nominal orbit from Table \ref{table:TLE}, space weather at that time, and dry mass and drag surface values for a \textit{Falcon 9} upper stage ($4500\, kg$ and $58.5\, m^{2}$ maximum drag surface). The model indicated that the nominal orbit derived from the state vector should see the rocket stage survive perigee and continue for a few additional revolutions.

However, an orbit that does result in deorbit over the designated area is possible within the error margins of the speed vector. A reduction in the speed vector of only $0.015\, km/s$ brings the perigee altitude of the orbit to below $80\, km$ and modeling in GMAT then results in a deorbit inside the designated zone from the \textit{Navigational Warning HYDROPAC 463/21}. Modeling was performed twice: once for a maximum drag surface of $58.5\, m^{2}$ and once for a minimum drag surface of $10.5\, m^{2}$. The rocket-stage dry mass used in both cases was 4500 kg. The resulting modeled impact points were near $38.0\, S$, $79.2\, E$ (maximum drag surface scenario), and $49.3\, S$, $104.3\, E$ (minimum drag surface scenario), which are both within the area indicated by the \textit{Navigational Warning HYDROPAC 463/21}. Furthermore, as indicated earlier, our state vector may actually represent the start of the deorbit burn, so it may not capture the full effect of the burn on the final reentry orbit, which could lead to an overestimation of the perigee altitude.

Based on these observations, practically no acceleration can be inferred. In the observed path, the velocity change ranges from $7.4499\, km/s$ to $7.4644\, km/s$ (from the proxy orbit). However, from the ground stations, it is difficult to obtain high-accuracy velocity measurements (below $0.1\, km/s$). We observed a velocity of $7.5 \pm 0.3\, km/s$, which does not allow us to appreciate the speed changes. Considering the fact that the recording contains the pre-ignition fuel cloud, subsequent engine burn, and consistency between the debris pieces orbits, the \textit{Falcon 9 Starlink V1.0-L19} launch and our observations are evidence that the SPMN160221ART event was a controlled deorbiting maneuver in a quasi-circular orbit.

\section{Discussion}
\noindent It seems clear that the increasing use of the near-Earth environment for commercial purposes will make reentries more frequent events (even with an increasing practice of deliberate deorbiting over the southern Pacific Ocean at the end of service life). In fact, another reentry was widely observed over the US on March 26 of 2021. In that case, the object was a \textit{Falcon 9} upper stage from from the March 4, 2021 \textit{Starlink} launch. The \textit{Falcon 9} upper stage failed to deorbit for unknown reasons and came down uncontrolled over Oregon and Washington in the northwest of the United States. This attracted significant public attention and several casual eyewitnesses filmed the event using their mobile phones. These recent events exemplify why we should be able to recognize and explain the real nature of such appearances. For the aforementioned reasons, it is relevant to develop a common methodology and software solution.

Regarding the event on February 16 of 2021, which is the subject of this study, we should discuss the sequence of events of typical controlled rocket stage reentries to understand our observations. A deorbit burn has at least two phases: an actual engine burn and propellant blow-out (fuel vent) at the end of the burn to avoid explosive disintegration of the rocket stage. Fuel vents (propellant blow-outs) often generate a circular cloud, sometimes (particularly with \textit{Falcon 9} stages, where it has been reported several times) with a spiral shape if the stage is rotating (spin stabilization). The fuel cloud initially moves at the same speed as the rocket stage from which it originates (the cloud co-orbits with the rocket stage). Over time, it expands, and differential drag and small $\Delta V$ differences separate it from the rocket stage. However, just after blow-out, it will stay with the rocket stage for a short duration as it begins radially expanding away from the rocket stage \cite{Oberg2018}. This is what we observed in our records of the February 16 event. In the early part of the Estepa North camera record, starting just after 5:53:45 UTC, a diffuse circular cloud can be observed surrounding one of two faint objects (see Figure \ref{fig:Fig1}. This is approximately 4 min before the start of the sudden bright phase. One of the two faint objects is likely the clump of released payloads. The other, which is centered in the diffuse cloud, is likely the Falcon 9 upper stage. This suggests that a burn or propellant blow-out (tank depressurization) occurred just before the start of camera recording. The beginning of visibility in the Estepa North camera record corresponds to the time at which rocket stage and generated fuel cloud passed from the Earth’s shadow into sunlight at approximately 5:53:45 UTC.

These findings make it very unlikely that the sudden bright phase appearing during the pass imaged by both Benicàssim and Estepa was caused by ablation. Therefore, we prefer the interpretation that the bright phase represents the actual deorbit engine burn. Alternatively, it could be generated by the payloads flaring up when the sun-payload-observer angle and the angles of the payload surfaces are favorable. The payloads may be very bright occasionally and in this phase of the mission, they are still close to the rocket stage (as shown at the start of visibility in the Estepa video). The pre-burn proxy orbit and the orbit derived from the state vector resulting from the observations converge to within $10\, km$ at 5:57:48 UTC. This is very close to the start of the sudden bright phase. The fuel cloud captured a few minutes earlier could be attributed to an earlier maneuver (e.g., a payload avoidance maneuver).

\section{Conclusions}
\noindent It is becoming increasingly common to observe luminous objects moving through the sky. Although most fireballs are natural and have a meteoric origin, a growing number of human-made objects in orbit around the Earth is leading to countless sightings and detection records of impressive light phenomena in the sky. In this work, we demonstrated how a traditional fireball analysis technique can be adapted to compute the curved trajectories of objects experiencing deorbit and reentry, as well as how to handle long recording durations. 

\begin{itemize}
\item We developed an extension for the \textit{3D-FireTOC Python} software by modifying the plane intersection method for meteor triangulation to be able to analyze slow objects with curved orbits captured over long periods of time.

\item We computed a \textit{Falcon 9} upper stage trajectory from video recordings obtained by two SPMN network stations on February 16 of 2021. The results were successfully compared to an orbit estimate from the orbital parameters published by CSpOC for four debris pieces associated with the \textit{Starlink V1.0-L19} launch. 

\item Our data confirmed that the recordings were obtained during or just after the deorbit burn of the rocket stage, meaning this type of pre-ablation phase in a reentry has the potential to produce eyewitness sightings causing varying degrees of alarm because such sightings are typically unexpected.

\item By using the \textit{3D-FireTOC} pipeline, the final orbit and deorbit trajectory of the \textit{Falcon 9} stage were successfully modeled. The results indicated a reentry over the area in the Indian Ocean that was designated for this entry, which is an additional added value of reentry tracking from the ground that increases our capacity to recover space junk arriving on the ground.

\item It seems clear that the advancement of space technology will lead to an increase in satellite sightings and artificial bolides produced by the reentry of space debris and rocket stages. This may interfere with astronomy in general and fireball studies in particular. 

\item Based on the large amount of data recorded and the proliferation of new detection stations around the world, fireball analysis processes are being automated. Therefore, the development of false-positive avoidance techniques to rule out unnatural events of no scientific interest is urgent. 

\item This type of event will become increasingly common and our results exemplify how ground-based fireball station networks could gradually play an important role in monitoring the deorbit of hazardous artificial objects in near-Earth space.

\end{itemize}

\subsection*{Acknowledgements}
\noindent This research was supported by the research project (PGC2018-097374-B-I00, PI: JMT-R), which is funded by FEDER/Ministerio de Ciencia e Innovación - Agencia Estatal de Investigación. This project has also received funding from the European Research Council (ERC) under the European Union’s Horizon 2020 Research and Innovation Programme (grant agreement No. 865657) for the project “Quantum Chemistry on Interstellar Grains” (QUANTUMGRAIN). We also express appreciation for the valuable video recordings obtained from Benicàssim (Castellón) by Vicent Ibáñez (AVAMET).\\

\refs{References}
{\zihao{5-} 
\bibliographystyle{astrobib} 
\bibliography{refs}           

\begin{thebibliography}{10}
\expandafter\ifx\csname urlstyle\endcsname\relax
  \providecommand{\doi}[1]{doi:\discretionary{}{}{}#1}\else
  \providecommand{\doi}{doi:\discretionary{}{}{}\begingroup
  \urlstyle{rm}\Url}\fi

\bibitem{ceplecha1998meteor}
Ceplecha Z, Borovi{\v{c}}ka J, Elford WG, ReVelle DO, Hawkes RL, Porub{\v{c}}an
  V, {\v{S}}imek M. Meteor phenomena and bodies. \emph{Space Science Reviews},
  1998, 84(3): 327--471.

\bibitem{opik1959distribution}
{{\"O}pik} EJ, {Singer} SF. {Distribution of Density in a Planetary Exosphere}.
  \emph{Physics of Fluids}, 1959, 2: 653--655, \doi{10.1063/1.1705968}.

\bibitem{revelle1979quasi}
{Revelle} DO. {A Quasi-Simple Ablation Model for Large Meteorite Entry: Theory
  VS Observations}. \emph{Journal of Atmospheric and Terrestrial Physics},
  1979, 41: 453--473, \doi{10.1016/0021-9169(79)90071-0}.

\bibitem{trigo2019flux}
Trigo-Rodr{\'\i}guez JM. The flux of meteoroids over time: meteor emission
  spectroscopy and the delivery of volatiles and chondritic materials to Earth.
  \emph{Hypersonic Meteoroid Entry Physics}, 2019: 4.

\bibitem{rubin2010meteorite}
{Rubin} AE, {Grossman} JN. {Meteorite and meteoroid: New comprehensive
  definitions}. \emph{MAstronomy and Astrophysics}, 2010, 45(1): 114--122,
  \doi{10.1111/j.1945-5100.2009.01009.x}.

\bibitem{jacchia1956harvard}
{Jacchia} LG, {Whipple} FL. {The Harvard photographic meteor programme}.
  \emph{Vistas in Astronomy}, 1956, 2(1): 982--994,
  \doi{10.1016/0083-6656(56)90021-6}.

\bibitem{ceplecha1957photographic}
{Ceplecha} Z. {Photographic Geminids 1955}. \emph{Bulletin of the Astronomical
  Institutes of Czechoslovakia}, 1957, 8: 51.

\bibitem{bland2004desert}
{Bland} PA. {Fireball cameras: The Desert Fireball Network}. \emph{Astronomy
  and Geophysics}, 2004, 45(5): 5.20--5.23,
  \doi{10.1046/j.1468-4004.2003.45520.x}.

\bibitem{trigo2005development}
{Trigo-Rodr{\'\i}guez} JM, {Castro-Tirado} AJ, {Llorca} J, {Fabregat} J,
  {Mart{\'\i}nez} VJ, {Reglero} V, {Jel{\'\i}nek} M, {Kub{\'a}nek} P, {Mateo}
  T, {de Ugarte Postigo} A. {The Development of the Spanish Fireball Network
  Using a New All-Sky CCD System}. \emph{Earth Moon and Planets}, 2004,
  95(1-4): 553--567, \doi{10.1007/s11038-005-4341-9}.

\bibitem{weryk2007southern}
{Weryk} RJ, {Brown} PG, {Domokos} A, {Edwards} WN, {Krzeminski} Z, {Nudds} SH,
  {Welch} DL. {The Southern Ontario All-sky Meteor Camera Network}. \emph{Earth
  Moon and Planets}, 2008, 102(1-4): 241--246, \doi{10.1007/s11038-007-9183-1}.

\bibitem{gritsevich2014first}
{Gritsevich} M, {Lyytinen} E, {Moilanen} J, {Kohout} T, {Dmitriev} V, {Lupovka}
  V, {Midtskogen} V, {Kruglikov} N, {Ischenko} A, {Yakovlev} G, {Grokhovsky} V,
  {Haloda} J, {Halodova} P, {Peltoniemi} J, {Aikkila} A, {Taavitsainen} A,
  {Lauanne} J, {Pekkola} M, {Kokko} P, {Lahtinen} P, {Larionov} M. {First
  meteorite recovery based on observations by the Finnish Fireball Network}. In
  JL~{Rault}, P~{Roggemans}, editors, \emph{Proceedings of the International
  Meteor Conference, Giron, France, 18-21 September 2014}, 2014, 162--169.

\bibitem{colas2015french}
{Colas} F, {Zanda} B, {Vaubaillon} J, {Bouley} S, {Marmo} C, {Audureau} Y,
  {Kwon} MK, {Rault} JL, {Caminade} S, {Vernazza} P, {Gattacceca} J, {Birlan}
  M, {Maquet} L, {Egal} A, {Rotaru} M, {Birnbaum} C, {Cochard} F, {Thizy} O.
  {French fireball network FRIPON}. In \emph{International Meteor Conference
  Mistelbach, Austria}, 2015, 37.

\bibitem{colas2020fripon}
{Colas} F, {Zanda} B, {Bouley} S, {Jeanne} S, {Malgoyre} A, {Birlan} M,
  {Blanpain} C, {Gattacceca} J, {Jorda} L, {Lecubin} J, {Marmo} C, {Rault} JL,
  {Vaubaillon} J, {Vernazza} P, {Yohia} C, {Gardiol} D, {Nedelcu} A, {Poppe} B,
  {Rowe} J, {Forcier} M, {Koschny} D, {Trigo-Rodriguez} JM, {Lamy} H, {Behrend}
  R, {Ferri{\`e}re} L, {Barghini} D, {Buzzoni} A, {Carbognani} A, {Di Carlo} M,
  {Di Martino} M, {Knapic} C, {Londero} E, {Pratesi} G, {Rasetti} S, {Riva} W,
  {Stirpe} GM, {Valsecchi} GB, {Volpicelli} CA, {Zorba} S, {Coward} D,
  {Drolshagen} E, {Drolshagen} G, {Hernandez} O, {Jehin} E, {Jobin} M, {King}
  A, {Nitschelm} C, {Ott} T, {Sanchez-Lavega} A, {Toni} A, {Abraham} P,
  {Affaticati} F, {Albani} M, {Andreis} A, {Andrieu} T, {Anghel} S, {Antaluca}
  E, {Antier} K, {App{\'e}r{\'e}} T, {Armand} A, {Ascione} G, {Audureau} Y,
  {Auxepaules} G, {Avoscan} T, {Baba Aissa} D, {Bacci} P, {B{\v{a}}descu} O,
  {Baldini} R, {Baldo} R, {Balestrero} A, {Baratoux} D, {Barbotin} E, {Bardy}
  M, {Basso} S, {Bautista} O, {Bayle} LD, {Beck} P, {Bellitto} R, {Belluso} R,
  {Benna} C, {Benammi} M, {Beneteau} E, {Benkhaldoun} Z, {Bergamini} P,
  {Bernardi} F, {Bertaina} ME, {Bessin} P, {Betti} L, {Bettonvil} F, {Bihel} D,
  {Birnbaum} C, {Blagoi} O, {Blouri} E, {Boac{\u{a}}} I, {Boat{\v{a}}} R,
  {Bobiet} B, {Bonino} R, {Boros} K, {Bouchet} E, {Borgeot} V, {Bouchez} E,
  {Boust} D, {Boudon} V, {Bouman} T, {Bourget} P, {Brandenburg} S, {Bramond} P,
  {Braun} E, {Bussi} A, {Cacault} P, {Caillier} B, {Calegaro} A, {Camargo} J,
  {Caminade} S, {Campana} APC, {Campbell-Burns} P, {Canal-Domingo} R, {Carell}
  O, {Carreau} S, {Cascone} E, {Cattaneo} C, {Cauhape} P, {Cavier} P,
  {Celestin} S, {Cellino} A, {Champenois} M, {Chennaoui Aoudjehane} H,
  {Chevrier} S, {Cholvy} P, {Chomier} L, {Christou} A, {Cricchio} D, {Coadou}
  P, {Cocaign} JY, {Cochard} F, {Cointin} S, {Colombi} E, {Colque Saavedra} JP,
  {Corp} L, {Costa} M, {Costard} F, {Cottier} M, {Cournoyer} P, {Coustal} E,
  {Cremonese} G, {Cristea} O, {Cuzon} JC, {D'Agostino} G, {Daiffallah} K,
  {D{\v{a}}nescu} C, {Dardon} A, {Dasse} T, {Davadan} C, {Debs} V, {Defaix} JP,
  {Deleflie} F, {D'Elia} M, {De Luca} P, {De Maria} P, {Deverch{\`e}re} P,
  {Devillepoix} H, {Dias} A, {Di Dato} A, {Di Luca} R, {Dominici} FM, {Drouard}
  A, {Dumont} JL, {Dupouy} P, {Duvignac} L, {Egal} A, {Erasmus} N, {Esseiva} N,
  {Ebel} A, {Eisengarten} B, {Federici} F, {Feral} S, {Ferrant} G, {Ferreol} E,
  {Finitzer} P, {Foucault} A, {Francois} P, {Fr{\^\i}ncu} M, {Froger} JL,
  {Gaborit} F, {Gagliarducci} V, {Galard} J, {Gardavot} A, {Garmier} M,
  {Garnung} M, {Gautier} B, {Gendre} B, {Gerard} D, {Gerardi} A, {Godet} JP,
  {Grandchamps} A, {Grouiez} B, {Groult} S, {Guidetti} D, {Giuli} G, {Hello} Y,
  {Henry} X, {Herbreteau} G, {Herpin} M, {Hewins} P, {Hillairet} JJ, {Horak} J,
  {Hueso} R, {Huet} E, {Huet} S, {Hyaum{\'e}} F, {Interrante} G, {Isselin} Y,
  {Jeangeorges} Y, {Janeux} P, {Jeanneret} P, {Jobse} K, {Jouin} S, {Jouvard}
  JM, {Joy} K, {Julien} JF, {Kacerek} R, {Kaire} M, {Kempf} M, {Koschny} D,
  {Krier} C, {Kwon} MK, {Lacassagne} L, {Lachat} D, {Lagain} A, {Laisn{\'e}} E,
  {Lanchares} V, {Laskar} J, {Lazzarin} M, {Leblanc} M, {Lebreton} JP,
  {Lecomte} J, {Le D{\^u}} P, {Lelong} F, {Lera} S, {Leoni} JF, {Le-Pichon} A,
  {Le-Poupon} P, {Leroy} A, {Leto} G, {Levansuu} A, {Lewin} E, {Lienard} A,
  {Licchelli} D, {Locatelli} H, {Loehle} S, {Loizeau} D, {Luciani} L, {Maignan}
  M, {Manca} F, {Mancuso} S, {Mandon} E, {Mangold} N, {Mannucci} F, {Maquet} L,
  {Marant} D, {Marchal} Y, {Marin} JL, {Martin-Brisset} JC, {Martin} D,
  {Mathieu} D, {Maury} A, {Mespoulet} N, {Meyer} F, {Meyer} JY, {Meza} E,
  {Moggi Cecchi} V, {Moiroud} JJ, {Millan} M, {Montesarchio} M, {Misiano} A,
  {Molinari} E, {Molau} S, {Monari} J, {Monflier} B, {Monkos} A, {Montemaggi}
  M, {Monti} G, {Moreau} R, {Morin} J, {Mourgues} R, {Mousis} O, {Nablanc} C,
  {Nastasi} A, {Niac{\c{s}}u} L, {Notez} P, {Ory} M, {Pace} E, {Paganelli} MA,
  {Pagola} A, {Pajuelo} M, {Palaci{\'a}n} JF, {Pallier} G, {Paraschiv} P,
  {Pardini} R, {Pavone} M, {Pavy} G, {Payen} G, {Pegoraro} A,
  {Pe{\~n}a-Asensio} E, {Perez} L, {P{\'e}rez-Hoyos} S, {Perlerin} V, {Peyrot}
  A, {Peth} F, {Pic} V, {Pietronave} S, {Pilger} C, {Piquel} M, {Pisanu} T,
  {Poppe} M, {Portois} L, {Prezeau} JF, {Pugno} N, {Quantin} C, {Quitt{\'e}} G,
  {Rambaux} N, {Ravier} E, {Repetti} U, {Ribas} S, {Richard} C, {Richard} D,
  {Rigoni} M, {Rivet} JP, {Rizzi} N, {Rochain} S, {Rojas} JF, {Romeo} M,
  {Rotaru} M, {Rotger} M, {Rougier} P, {Rousselot} P, {Rousset} J, {Rousseu} D,
  {Rubiera} O, {Rudawska} R, {Rudelle} J, {Ruguet} JP, {Russo} P, {Sales} S,
  {Sauzereau} O, {Salvati} F, {Schieffer} M, {Schreiner} D, {Scribano} Y,
  {Selvestrel} D, {Serra} R, {Shengold} L, {Shuttleworth} A, {Smareglia} R,
  {Sohy} S, {Soldi} M, {Stanga} R, {Steinhausser} A, {Strafella} F, {Sylla
  Mbaye} S, {Smedley} ARD, {Tagger} M, {Tanga} P, {Taricco} C, {Teng} JP,
  {Tercu} JO, {Thizy} O, {Thomas} JP, {Tombelli} M, {Trangosi} R, {Tregon} B,
  {Trivero} P, {Tukkers} A, {Turcu} V, {Umbriaco} G, {Unda-Sanzana} E,
  {Vairetti} R, {Valenzuela} M, {Valente} G, {Varennes} G, {Vauclair} S,
  {Vergne} J, {Verlinden} M, {Vidal-Alaiz} M, {Vieira-Martins} R, {Viel} A,
  {V{\^\i}ntdevar{\v{a}}} DC, {Vinogradoff} V, {Volpini} P, {Wendling} M,
  {Wilhelm} P, {Wohlgemuth} K, {Yanguas} P, {Zagarella} R, {Zollo} A. {FRIPON:
  a worldwide network to track incoming meteoroids}. \emph{Astronomy and
  Astrophysics}, 2020, 644: A53, \doi{10.1051/0004-6361/202038649}.

\bibitem{gardiol2016prisma}
{Gardiol} D, {Cellino} A, {Di Martino} M. {{PRISMA}, Italian network for
  meteors and atmospheric studies}. In A~{Roggemans}, P~{Roggemans}, editors,
  \emph{International Meteor Conference Egmond, the Netherlands, 2-5 June
  2016}, 2016, 76.

\bibitem{devillepoix2020global}
{Devillepoix} HAR, {Cup{\'a}k} M, {Bland} PA, {Sansom} EK, {Towner} MC, {Howie}
  RM, {Hartig} BAD, {Jansen-Sturgeon} T, {Shober} PM, {Anderson} SL, {Benedix}
  GK, {Busan} D, {Sayers} R, {Jenniskens} P, {Albers} J, {Herd} CDK, {Hill}
  PJA, {Brown} PG, {Krzeminski} Z, {Osinski} GR, {Aoudjehane} HC, {Benkhaldoun}
  Z, {Jabiri} A, {Guennoun} M, {Barka} A, {Darhmaoui} H, {Daly} L, {Collins}
  GS, {McMullan} S, {Suttle} MD, {Ireland} T, {Bonning} G, {Baeza} L, {Alrefay}
  TY, {Horner} J, {Swindle} TD, {Hergenrother} CW, {Fries} MD, {Tomkins} A,
  {Langendam} A, {Rushmer} T, {O'Neill} C, {Janches} D, {Hormaechea} JL, {Shaw}
  C, {Young} JS, {Alexander} M, {Mardon} AD, {Tate} JR. {A Global Fireball
  Observatory}. \emph{Planetary and Space Science}, 2020, 191: 105036,
  \doi{10.1016/j.pss.2020.105036}.

\bibitem{trigo20072006}
{Trigo-Rodr{\'\i}guez} JM, {Madiedo} JM, {Llorca} J, {Gural} PS, {Pujols} P,
  {Tezel} T. {The 2006 Orionid outburst imaged by all-sky CCD cameras from
  Spain: meteoroid spatial fluxes and orbital elements}. \emph{Monthly Notices
  of the Royal Astronomical Society}, 2007, 380(1): 126--132,
  \doi{10.1111/j.1365-2966.2007.11966.x}.

\bibitem{madiedo2007multi}
{Madiedo} JM, {Trigo-Rodr{\'\i}guez} JM. {Multi-station Video Orbits of Minor
  Meteor Showers}. \emph{Earth Moon and Planets}, 2008, 102(1-4): 133--139,
  \doi{10.1007/s11038-007-9215-x}.

\bibitem{klinkrad2010space}
Klinkrad H. Space Debris: Models and Risk Analysis, 2006,
  \doi{10.1007/3-540-37674-7}.

\bibitem{liou2006risks}
Liou J, Johnson NL. Risks in space from orbiting debris. \emph{Science-New York
  Then Washington-}, 2006, 311(5759): 340.

\bibitem{hainaut2020impact}
{Hainaut} OR, {Williams} AP. {Impact of satellite constellations on
  astronomical observations with ESO telescopes in the visible and infrared
  domains}. \emph{Astronomy and Astrophysics}, 2020, 636: A121,
  \doi{10.1051/0004-6361/202037501}.

\bibitem{bagrov2010calculation}
{Bagrov} AV, {Leonov} VA. {The calculation of meteor motion parameters based on
  the single station TV observations}. \emph{Solar System Research}, 2010,
  44(4): 327--333, \doi{10.1134/S0038094610040064}.

\bibitem{mironov2021retrospective}
{Mironov} VV, {Murtazov} AK. {Retrospective on the Problem of Space Debris.
  Part 2. Monitoring of Space Debris of Natural Origin in Near-Earth Space
  Using Optical Methods of Meteor Astronomy}. \emph{Cosmic Research}, 2021,
  59(1): 36--45, \doi{10.1134/S0010952521010056}.

\bibitem{revelle2005genesis}
{Revelle} DO, {Edwards} W, {Sandoval} TD. {Genesis--An artificial, low velocity
  ``meteor'' fall and recovery: September 8, 2004}. \emph{Meteoritics and
  Planetary Science}, 2005, 40: 895, \doi{10.1111/j.1945-5100.2005.tb00162.x}.

\bibitem{revelle2007stardust}
{Revelle} DO, {Edwards} WN. {Stardust{\textemdash}An artificial, low-velocity
  ``meteor'' fall and recovery: 15 January 2006}. \emph{Meteoritics and
  Planetary Science}, 2007, 42(2): 271--299,
  \doi{10.1111/j.1945-5100.2007.tb00232.x}.

\bibitem{levit2008reconstruction}
Levit C, Albers J, Jenniskens P, Spurny P. Reconstruction and Verification of
  the {Stardust SRC} Re-Entry Trajectory. \emph{AIAA paper}, 2008: 1199.

\bibitem{pas2009atv}
de~Pasquale E, Francillout L, Wasbauer JJ, Hatton J, Albers J, Steele D. ATV
  Jules Verne reentry observation: Mission design and trajectory analysis. In
  \emph{2009 IEEE Aerospace conference}, 2009, 1--16,
  \doi{10.1109/AERO.2009.4839703}.

\bibitem{ueda2011trajectory}
{Ueda} M, {Shiba} Y, {Yamamoto} My, {Fujita} K, {Watanabe} Ji, {Sato} M, {Abe}
  S, {Kakinami} Y, {Uehara} S, {Okamoto} S, {Fujiwara} Y, {Tanabe} T.
  {Trajectory of HAYABUSA Reentry Determined from Multisite TV Observations}.
  \emph{Publications of the Astronomical Society of Japan}, 2011, 63(5):
  947--953, \doi{10.1093/pasj/63.5.947}.

\bibitem{shoemaker2013trajectory}
{Shoemaker} MA, {van der Ha} JC, {Abe} S, {Fujita} K. {Trajectory Estimation of
  the Hayabusa Spacecraft During Atmospheric Disintegration}. \emph{Journal of
  Spacecraft and Rockets}, 2013, 50(2): 326--336, \doi{10.2514/1.A32338}.

\bibitem{pena2021accurate}
{Pe{\~n}a-Asensio} E, {Trigo-Rodr{\'\i}guez} JM, {Gritsevich} M, {Rimola} A.
  {Accurate 3D fireball trajectory and orbit calculation using the 3D-FIRETOC
  automatic Python code}. volume 504, 2021, 4829--4840,
  \doi{10.1093/mnras/stab999}.

\bibitem{ceplecha1987geometric}
{Ceplecha} Z. {Geometric, Dynamic, Orbital and Photometric Data on Meteoroids
  From Photographic Fireball Networks}. \emph{Bulletin of the Astronomical
  Institutes of Czechoslovakia}, 1987, 38: 222.

\bibitem{borovivcka1992astrometry}
{Borovi{\v{c}}ka} J. {Astrometry with all-sky cameras.} \emph{Publications of
  the Astronomical Institute of the Czechoslovak Academy of Sciences}, 1992,
  79.

\bibitem{borovicka1995new}
{Borovicka} J, {Spurny} P, {Keclikova} J. {A new positional astrometric method
  for all-sky cameras.} \emph{Astronomy and Astrophysics Supplement}, 1995,
  112: 173.

\bibitem{bannister2013numerical}
{Bannister} SM, {Boucheron} LE, {Voelz} DG. {A Numerical Analysis of a Frame
  Calibration Method for Video-based All-Sky Camera Systems}. \emph{Monthly
  Notices of the Royal Astronomical Society}, 2013, 125(931): 1108,
  \doi{10.1086/673167}.

\bibitem{motzkin1956assignment}
Motzkin T. The assignment problem. In \emph{Proc. Symposia in Applied
  Mathematics}, volume~6, 1956, 109--125.

\bibitem{pena2020}
Pe{\~n}a-Asensio E, {Trigo-Rodr{\'\i}guez} JM, {Mas-Sanz} E, {Ribas} J.
  {SPMN160819 superbolide: reconstructing its atmospheric trajectory by
  matching ground-based recordings and satellite data}. In \emph{European
  Planetary Science Congress}, 2020, EPSC2020--459.

\bibitem{dubiago1961determination}
{Dubiago} AD. \emph{{The determination of orbits.}} 1961.

\bibitem{Hoots1980}
Hoots FR, Roehrich RL. Models for Propagation of {NORAD} Element Sets.
  Technical report, 1980, \doi{10.21236/ada093554}.

\bibitem{Vallado2006}
Vallado D, Crawford P, Hujsak R, Kelso T. Revisiting Spacetrack Report {\#}3,
  2006, \doi{10.2514/6.2006-6753}.

\bibitem{osweiler2006covariance}
Osweiler VP. Covariance estimation and autocorrelation of NORAD two-line
  element sets. Ph.D. thesis, US Air Force Institute of Technology, 2006.

\bibitem{kelso2007validation}
Kelso T, et~al.. Validation of SGP4 and IS-GPS-200D against GPS precision
  ephemerides. In \emph{17th AAS/AIAA Space Flight Mechanics Conference Paper
  AAS 07-127}, 2007.

\bibitem{Moriba2009}
{Jah} M, {Huges} S, {Wilkins} M, {Kelecy} T. {The General Mission Analysis Tool
  (GMAT): A New Resource for Supporting Debris Orbit Determination, Tracking
  and Analysis}. In H~{Lacoste}, editor, \emph{Fifth European Conference on
  Space Debris}, volume 672 of \emph{ESA Special Publication}, 2009, 124.

\bibitem{Oberg2018}
Oberg J. Ground observations of {Falcon-9} second stage orbital
  venting/thrusting as aid for interpreting unusual visual features of
  mysterious {‘Zuma’} launch, 2018.

\end{thebibliography}
}

\subsection*{Authors biography} 
\label{subsec:author}

\begin{wrapfigure}{L}{0.12\textwidth}
\centering
\includegraphics[width=1in,height=1.25in,clip,keepaspectratio]{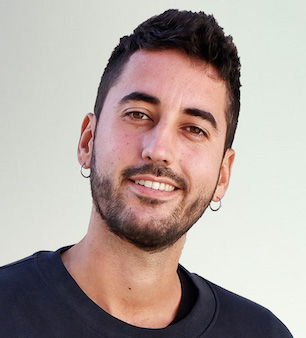}
\end{wrapfigure}

Eloy Peña Asensio was born in Cartagena, Spain. He holds a degree in aerospace engineering from the Technical University of Madrid (UPM), a Master’s degree in Aerospace Science and Technology from the Polytechnic University of Catalonia (UPC), and a Master’s degree in astrophysics and cosmology from the Autonomous University of Barcelona (UAB). He has two years of experience at the Institute for Space Studies of Catalonia (IEEC) developing a system for the automatic detection and analysis of meteors and large bolides for the Spanish Meteor Network (SPMN). In 2018, he performed an internship at the Manipal Institute of Technology in India for three months while working on the system identification of twin rotor systems. He was also working to develop a drone system for drifting boat sightings using computer vision for the NGO Open Arms in conjunction with the HEMAV foundation in 2019. He is currently working on his PhD thesis on meteorites under the supervision of Albert Rimola and Josep M. Trigo-Rodríguez at UAB in the context of the European ERC QUANTUMGRAIN project. He was awarded by the Aerospace and Electronics Systems Society (AESS) of the IEEE for the best national Master’s thesis. He was recognized by the city council of his hometown as an extraordinary youth in 2021. \\

\begin{wrapfigure}{L}{0.12\textwidth}
\centering
\includegraphics[width=1in,height=1.25in,clip,keepaspectratio]{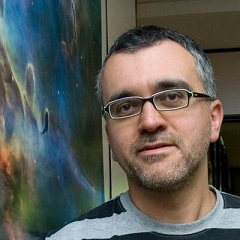}
\end{wrapfigure}
Josep M. Trigo-Rodriguez was born in Valencia, Spain, on July 3, 1970. He obtained his degree in physics at the University of Valencia in 1997 and his PhD in theoretical physics (astrophysics) in 2002 under the direction of Professor Jordi Llorca (UPC) and Professor Juan Fabregat (UV). He was a visitor at the Ondrejov Observatory during his pre-doctoral stay. In 2003, he received a USA-Spanish grant that allowed him to continue his career in a postdoctoral position at the Institute of Geophysics \& Planetary Physics of the University of California Los Angeles (UCLA) and the NASA Astrobiology Center at UCLA under the supervision of Professor John Wasson and Doctor Alan Rubin. After almost three years of work on the transport of water and volatiles in primitive meteorites (carbonaceous chondrites), he returned to Spain in 2006 with a Juan de la Cierva grant to join the Institute of Space Sciences (ICE, CSIC-IEEC) in Barcelona, Catalonia. In 2009, he received his position as a Tenured Scientist of the Upper Research Council (CSIC) at the same research institute. Since 2010, Dr. Trigo-Rodriguez has been the leader of the Meteorite, Minor Bodies, and Planetary Sciences Group at ICE (CSIC-IEEC). His current research focuses on the formation of primitive solar system minor bodies (comets and asteroids), the study of their fragments in space (dust, meteoroids), and the analysis and characterization of surviving rocks arriving on the Earth (meteorites). These “minor bodies” provide clues regarding the origin of the solar system because are retentive of the protoplanetary disc components that contain clues regarding the chemical and isotopic conditions prevailing in the early solar system. Since 2012, Dr. Trigo-Rodríguez has taught astrobiology, astrophysics, and planetary sciences in two international Master’s programs: MasterCosmosBCN (postgraduate program of High Energy Physics, Astrophysics \& Cosmology) and the Valencian International University (VIU). He has written 15 astronomy books in Catalan, English, and Spanish, and has received several awards for his scientific career and outreach tasks. Dr. Trigo-Rodríguez is Chief Editor of Advances in Astronomy and Associate Editor of three journals: Galaxies, Meteoritics and Planetary Science, and Frontiers in Astronomy and Space Science. He is also an editor of Impact Studies, which is a collection of books dedicated to impact hazards published by Springer.\\

\begin{wrapfigure}{L}{0.12\textwidth}
\centering
\includegraphics[width=1in,height=1.25in,clip,keepaspectratio]{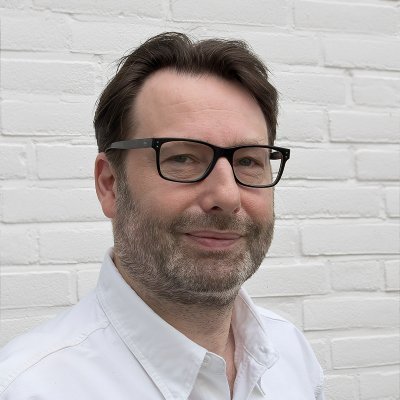}
\end{wrapfigure}
Marco Langbroek (1970) is a multidisciplinary scientist who studied prehistoric archeology at Leiden University in the Netherlands. He obtained his PhD in Paleolithic archeology at Leiden University in 2003 under the supervision of Professor Wil Roebroeks. He subsequently branched into other fields of science. These include asteroid discovery, meteor and meteorite research, and space situational awareness (SSA). He is a well-known tracker and analyst of classified military satellites. He has worked as an academic researcher among the Faculty of Archaeology at Leiden University, at the Institute for Geo- and Bioarchaeology (IGBA) at the VU University Amsterdam, and at the Department of Geology at the Naturalis Biodiversity Center in Leiden (the former Dutch National Museum of Natural History). He is currently working in the Department of Astronomy of Leiden University. From 2008 to 2012, with funding from a VENI grant from the Dutch National Science Foundation NWO, he studied the spatial behavior and cognition of Neandertals at the VU University Amsterdam. From 2012 to 2019, while working at VU University and later at Naturalis, he was the PI of the Diepenveen Meteorite Research Consortium. He and a large international team of co-workers published a study on the unique Dutch diepenveen CM-an carbonaceous chondrite. In the Astronomy Department of Leiden University, he currently works as a consultant on space situational awareness issues in the SOT project of Leiden University with the Space Security Center of the Royal Dutch Air Force. He is still affiliated as a guest researcher at the Naturalis Biodiversity Center. He received the Van Es Prize for Dutch Archeology in 1998 and the Doctor J. van der Bilt Prize of the Royal Dutch Association for Meteorology and Astronomy (KNVWS) in 2012. In 2008, the IAU named the asteroid (183294) Langbroek in his honor. He is active as a popular science educator, including appearances in news media and on Dutch radio and television on topics such as meteorites, fireballs, and satellites.\\

\begin{wrapfigure}{L}{0.12\textwidth}
\centering
\includegraphics[width=1in,height=1.25in,clip,keepaspectratio]{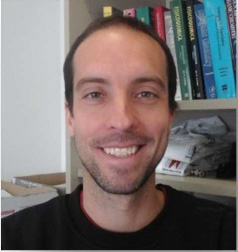}
\end{wrapfigure}
Albert Rimola graduated with a degree in chemistry from the Universitat Autònoma de Barcelona (UAB, 2002) and received a PhD in theoretical and computational chemistry (UAB, 2007) under the supervision of Professor M. Sodupe. He then conducted post-doctoral work (2007 to 2009) in a group under Professor Piero Ugliengo (Univ. Turin) and in 2010, he returned to the UAB. During these years, he obtained several grants and contracts through competitive international calls. He is currently a Ramón y Cajal researcher at UAB in a prestigious five year tenure-track position. His research focuses on the simulation of chemical processes through accurate quantum chemical calculations using both molecular and periodic ab initio approaches. His thesis focused on the interactions of open-shell transition metal cations with probe biomolecules by combining quantum chemical calculations and mass spectrometry experiments, which were linked to investigate the role of metal cations in Alzheimer’s disease. He acquired extensive knowledge of various quantum chemical methods and deep expertise in the simulation of chemical reactivity. During his post-doctoral research, he studied the electronic structures of different solid-state periodic systems and their adsorptive and chemical reactivity properties, acquiring extensive experience in surface modeling. His main expertise is in the simulation of chemical reactivity and modeling of solid-state surfaces, and his current research activities merge and exploit these two skills, which are of great importance in the field of grain surface chemistry.\\

\begin{wrapfigure}{L}{0.12\textwidth}
\centering
\includegraphics[width=1in,height=1.25in,clip,keepaspectratio]{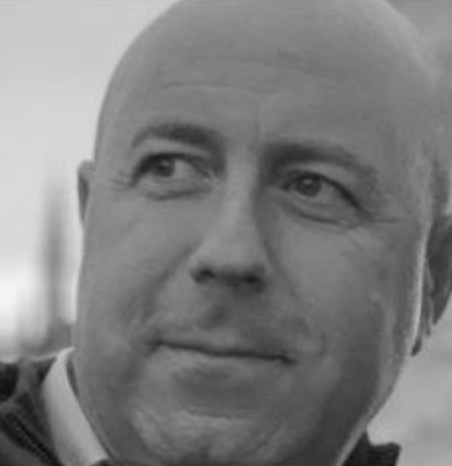}
\end{wrapfigure}
Antonio J. Robles was born in Seville in 1971. He studied architecture at the Escuela Técnica Superior de Arquitectura de Sevilla. He has been a member of the College of Architects of Seville since 2003. He is passionate about astronomy and has been collaborating as an amateur with the Spanish Meteor Network (SPMN). He has contributed significantly to the increase in large fireballs registered in the Iberian Peninsula from his detection station in Estepa, Seville.\\

\end{document}